\documentclass[twocolumn,superscriptaddress,showpacs,preprintnumbers,amsmath,amssymb]{revtex4}

\newcommand{\bq}{\begin{equation}}
\newcommand{\ba}{\begin{eqnarray}}
\newcommand{\eq}{\end{equation}}
\newcommand{\ea}{\end{eqnarray}}

\newcommand {\calD} {{\cal D}}

\usepackage{amsfonts}
\usepackage{amssymb}
\usepackage{amsmath}
\usepackage{feynmp}
\usepackage{color}
\usepackage{graphicx}
\usepackage[pdftex,colorlin ks=true]{hyperref}

\begin{document}

\title{Composite Bound States and Broken $U(1)$ symmetry in the Chemical Master Equation derivation of the Gray-Scott Model}
\author{Fred Cooper} \email{fcooper@fas.harvard.edu}
\affiliation{Department of Earth and Planetary Sciences, Harvard University, Cambridge, MA 02138}
\affiliation{The Santa Fe Institute, 1399 Hyde Park Road, Santa Fe, NM 87501, USA}
\author{Gourab Ghoshal}   \email{gghoshal@fas.harvard.edu}
\affiliation{Department of Earth and Planetary Sciences, Harvard University, Cambridge, MA 02138}
\author{Juan Perez-Mercader} \email{jperezmercader@fas.harvard.edu}
\affiliation{Department of Earth and Planetary Sciences, Harvard University, Cambridge, MA 02138}
\affiliation{The Santa Fe Institute, 1399 Hyde Park Road, Santa Fe, NM 87501, USA}

\date{\today}

\begin{abstract}

We give a first principles derivation  of the stochastic partial differential equations that describe the chemical reactions of the Gray-Scott model (GS):  $U+2V  {\stackrel {\lambda}{\rightarrow}}~ 3 V;$ and $V {\stackrel {\mu}{\rightarrow}}~P$, $U  {\stackrel {\nu}{\rightarrow}}~ Q$,  with a constant feed rate for $U$.  We find that the conservation of probability ensured by the chemical master equation leads to a modification of the usual differential equations for the GS model  which now involves  two composite fields and also intrinsic noise terms. One of the composites is  $\psi_1 = \phi_v^2$, where $\langle \phi_v \rangle_{\eta} = v$ is the concentration of the species $V$ and the averaging is over the internal noise $\eta_{u,v,\psi_1}$. The second composite field is the  product of three fields $ \chi =  \lambda \phi_u \phi_v^2$ and requires a noise source to ensure probability conservation.  A third composite $\psi_2 = \phi_{u} \phi_{v}$ can be also be identified from the noise-induced reactions. The Hamiltonian that governs the time evolution of the  many-body wave function, associated with the master equation, has a broken $U(1)$ symmetry related to particle number conservation. By expanding around the (broken symmetry) zero energy solution of the Hamiltonian (by performing a Doi shift)  one obtains from our path integral formulation the usual reaction diffusion equation, at the classical level. The Langevin equations that are derived from the chemical master equation have multiplicative noise sources for the  density fields $\phi_u, \phi_v, \chi$ that induce higher order processes such as $n \rightarrow n$ scattering for $n > 3$. The amplitude of the noise acting on $ \phi_v$ is itself stochastic in nature.  
\end{abstract}
\maketitle

\section{Introduction}
The dynamics of many systems in the real world are often the result of the interplay between their myriad constituents. Their bulk properties emerge as a manifestation of collective behavior and multiple subsystems where ``the whole is greater than the (simple) sum of its parts". Systems with these properties are broadly classified as emergent or complex systems \cite{Mandelbrot, PerBak, Holland}.  They are open systems subject to various kinds of fluctuations and the exchange of energy and/or matter with their environment.  The fluctuations take place both at the level of the whole system as well as at the scale of the interactions between the internal components. In particular, the fluctuations at the microscopic level arise from processes due to local collisions between molecules of different species, or to their interactions with the external environment.
The fact that in many of these systems one finds different dynamics at varying scales, suggests the presence of hierarchically organized subsystems obeying \emph{effective} dynamics that operate independently of each other~\cite{prlus} while remaining connected through some network. 

There are well-known examples of the above across a wide variety of systems. Among them, Reaction-Diffusion (RD) systems provide a versatile class of models (in many cases the effective result of coarse-grained versions of some complex chemistry) capable of capturing many qualitative features of chemical and biological systems~\cite{CG09, Walgraef97}. In this context the Gray-Scott (GS) model, consisting of two chemical species interacting at large scales, is of particular interest~\cite{GS83} as it displays a variety of spatio-temporal patterns~\cite{Pearson93} that include a form of self-replication mimicking the behavior found in bacterial cells.  When \emph{external} noise is included in the model---as stirring effects for example---it can be seen to affect transitions between patterns or even driving the system into modes where it ``adapts" to its local environment~\cite{LHMPM03,HLMPM03}.  

However, there is also some form of \emph{intrinsic} noise in the system whose effects are less understood. Qualitatively, one can interpret this intrinsic noise as a signature of the underlying mechanisms, or resulting ``fine-grained'' subsystems, that lead to the observed behavior of the GS model at the level of its chemical kinetics. Because of this, it is important to understand the precise nature and effects of this noise, as it is a harbinger for the internal structure of the system and plays a central role in the solution of the ``inverse" problem. That is, one is motivated to ask ``what is the internal dynamics consistent with the external dynamics''?  In this paper we make progress towards answering this question by developing and applying a strategy that uses some of the  tools of non-equilibrium field theory  \cite{Kamenev11, Altland}.

The GS Model involves two species $U$ and $V$ that undergo the chemical reactions:
\bq  
U+2V  {\stackrel {\lambda}{\rightarrow}}~ 3 V, ~~
V {\stackrel {\mu}{\rightarrow}}~P, ~~
U  {\stackrel {\nu}{\rightarrow}}~ Q,~~ {\stackrel {f}{\rightarrow}}~U.   
\label{reactions}
\eq
There is a cubic autocatalytic step for V at rate $\lambda$, and decay reactions at rates $\mu, \nu$ that transform $V$ and $U$ into inert products $P$ and $Q$. Finally, $U$ is fed into the system at a rate $f$. The phenomenological approach to study the dynamics of such systems utilizes the law of mass action and allows us to interpret the chemical reaction $U+2V {\stackrel {\lambda}{\rightarrow}} 3V$ as having the terms $\pm \lambda u v^2$ (where lower-case refers to concentrations) in its reaction kinetics. Following this and including diffusion as a first approximation to molecular motion, the equations that describe the kinetics of the system are
\ba 
\frac{\partial v}{\partial t} && = \lambda u v^2 - \mu v + D_v \nabla^2 v,  \nonumber \\
\frac{\partial u}{\partial t} && =f - \lambda u v^2 - \nu u + D_u \nabla^2 u.
 \label{NGS} 
\ea 
Here $u(\mathbf{x},t)$ and $v(\mathbf{x},t)$ represent the concentrations of the chemical species $U$ and $V$ and, as such, are fields in a $d+1$ dimensional space $(\mathbf x, t$), while $D_v, D_u$ are the diffusion constants for species $V,U$. 

\begin{figure*}[t!]
\centering 
\includegraphics[width=0.75\textwidth]{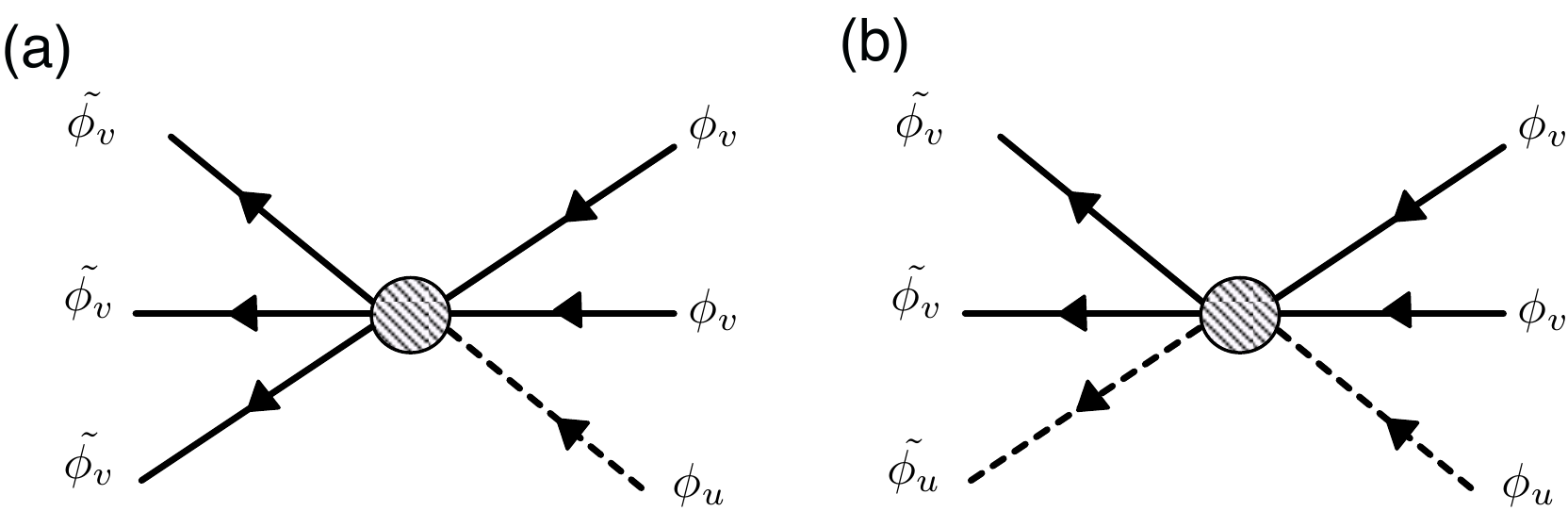}
\caption{Many body interpretation of the reaction $U+2V \rightarrow 3V$. (a) Inelastic scattering where one molecule of $U$ and two molecules of $V$ are destroyed at an interaction vertex (shaded) to create three molecules of $V$. (b) The elastic scattering where $U+2V \rightarrow U+2V$ which must be present in order for particle number and probability conservation. Time flows from left to right.}
\label{fig:fig1}
\end{figure*}

The set of equations in~\eqref{NGS} relies on the fundamental assumption of a separation of scales between the macroscopic dynamics (represented by their chemical kinetics) and internal dynamical processes at the microscopic level, where reactions proceed as collisions between individual molecules with probability proportional to the number of molecules of each species. 
Strictly speaking, this separation is valid only in the fully mixed regime which corresponds to either a well-stirred reactor or large diffusion coefficients, and ignores the spontaneous deviation from average behavior generated by fluctuations in the system.  For example, in the autocatalytic step in Eq.~\eqref{reactions} it is unlikely that the reaction takes place as the simultaneous local collision of well-mixed concentrations of three molecules. 

Of course, fluctuations are always present in macroscopic systems  due to the complexity of molecular--level motions which can be interpreted as generators of \emph{internal} noise for each species.  This then leads to the addition of noise terms in Eq.~\eqref{NGS} meant to represent deviations of the dynamics from the mean field regime. However, the introduction of these internal noise terms must not ``upset the apple cart".  That is, these noises must be such that in the mean field limit we recover the dynamics of the GS model~\eqref{NGS}. In the process of determining the noise terms we find that, in order for this condition to be satisfied we require the introduction of composite structures (which in fact, can be directly inferred by looking at the graphical structure of the scattering processes induced by the fluctuations). The existence of these structures suggests that the underlying dynamics must go through intermediate interactions occurring at shorter scales. 

A standard method of introducing collective variables in field theory is the use of functional Dirac delta functions or  the Hubbard-Stratonovich transformation \cite{Altland}.  By judicious use of the above methods we can introduce internal dynamical degrees of freedom as a result of promoting averages of two locally interacting fields into a single collective entity representing the larger scale degrees of freedom~\footnote{This  is the basis for establishing the well-known equivalence between the Ising model and $\lambda \phi^4$ field theory. See Ref. \cite{Amit, Binney, Altland}}.  

Following this,  we formulate an internal or ``fine-grained" dynamics which gives rise at the larger space-time scales to the fluctuations accompanying the chemistry in Eq.~\eqref{reactions}. In order to do so,  we employ a consistent non-equilibrium field theory in the form of a Langevin equation that captures the symmetries in the system as well as the underlying collective dynamics.

We first study the chemistry represented in Eq.  \eqref{reactions} and ask the question: what is the overall conserved probability distribution which describes the reactions at the level of molecular collisions? As is well known, this is answered by writing down the \emph{chemical master equation} for~\eqref{reactions} which, to first approximation provides a mechanistic description at this level~\cite{VK83}.  
We then use the formalism due to Doi~\cite{Doi76} to introduce an equivalent many body wave function $|\Psi \rangle $ that evolves according to a Hamiltonian $H$  that conserves probability.  In the many-body approach the reaction $U+ 2V \rightarrow 3V$ is described by an interaction vertex where one destroys one molecule of U and two molecules of V and then creates three molecules of V (see Fig.~\ref{fig:fig1}). This is a local interaction of six fields that involves three creation operators. Using this formalism, we find that the Hamiltonian has an explicitly broken $U(1)$ symmetry (corresponding to particle number conservation). Not surprisingly, in the particle conserving version of the Hamiltonian the  original $3\rightarrow3$ interaction gets modified by fluctuations that can be interpreted as the production and decay of a 3-body intermediate state.  These fluctuations in turn induce higher order particle number conserving interactions of the form $n \rightarrow n$ (where $n > 3$), which can be interpreted as ``tree graphs" in a theory with composite
fields made of various combinations of the chemicals $u, v$.  On the other hand,  the chemistry represented by Eq.~\eqref{NGS} arises in the ``broken symmetry" vacuum, corresponding to the zero energy point of the Hamiltonian. In this ``Doi-shifted''  Hamiltonian (which we will define below) one can identify the field associated with the annihilation operator directly with the chemical concentration.

Next we give a path integral description of the evolution operator $e^{-Ht}$ for both the shifted and un-shifted versions of $H$. In doing so we find terms in the action that are trilinear in the auxiliary fields $\phi^{\star} _{v,u}$. In order for us to derive the full Langevin equation describing the GS model with \emph{internally} generated noise we convert these trilinears to bilinears via the introduction of composite field operators.

This comes at the price of having to introduce additional terms. In fact, we will show that these terms in the modified Hamiltonian (which has only two creation operators) are generated by introducing noise terms $\eta_{v,u} (\mathbf x,t)$ for the fields  $\phi_u, \phi_v$ as well as $\eta_{\psi_1}(\mathbf x, t)$ for the composite field $\sigma = \lambda \phi_u \phi_v^2$. The composite field $\psi_1 = \phi_{v}^2$ induces the tree-level interactions $\sigma \rightarrow \phi_u + \psi_1$ and $\sigma \rightleftharpoons \phi_v + \psi_1$.  We find that the introduction of these composite fields is \emph{strictly} necessary for a Langevin description and these terms arise from satisfying probability conservation,  which underscores their important role in the internal dynamics of the GS model. 

In what follows we will use $\phi_{u,v}$ to denote the many-body fields and $\langle \phi_u \rangle_{\eta} =u, \langle \phi_v \rangle_{\eta} = v$ as the noise-averages of these fields. 
The composite field $\sigma$ is similar to the Hubbard--Stratonovich fields that enter in the discussion of large--$N$ expansions ~\cite{CJP74},  BCS theory (Cooper Pairs) \cite{dMRE93}  and BEC theory \cite{CCMD10}.  
The inverse propagator for the $\sigma$ field does not have a bare kinetic term (instead one has a  delta function in coordinate space)  as is appropriate for a composite field operator. The Langevin equations we derive are appropriate to the large scale dynamics that represent the chemistry given in~\eqref{reactions}, contains the inherent  symmetries present in the system and include the probability conserving fluctuations due to the Markovian nature of the interactions at the molecular level. 

We end the paper by summarizing our conclusions and presenting a map of further work to be done.

\section{Chemical master equation and many body formalism}
In order to develop the master equation formalism, we first divide the space in which the reactions take place into a $d-$dimensional hyper-cubic lattice of cells and assume that we can treat each cell as a coherent entity.  To do so, the system must satisfy the requirement of being at local mechanical and thermal equilibrium. In addition, the size of each cell must at least be of the order of the mean free path so that we can neglect microscopic attributes such as the velocity distribution functions of each molecule, and describe the state of the system completely in terms of its composition variables. The composition of the system changes through inelastic reactive collections, which are typically rare when compared to elastic non-reactive ones. This in turn implies that the evolution of the system can be represented by a \emph{jump Markov process} due to randomization effects by the elastic collisions.

Let  $\mathbf {n_i(t)} =( \{n_i(t) \}) $ be a vector composition variable where $n_i$  represents the number of molecules of a species at site $i$. Denoting $P(\mathbf{n_v, n_u} ,t)$ as the probability to find the particle configuration $\mathbf{(n_v, n_u)}$ at time $t$,  one obtains for the master equation~\cite{HZM_2005} for the chemical reactions in~\eqref{reactions} including diffusion 
\begin{widetext}
\ba \label{master2}
\frac{d}{dt}P(\mathbf{n_v, n_u},t) &=& \frac{D_v}{l^2}  \sum_{\langle i,j \rangle} \left[ (n_{v,j}+1)P(\ldots,n_{v,i}-1, n_{v,j}+1, \ldots, t) - n_{v,i}P \right] \nonumber \\
&+& \frac{D_u}{l^2}  \sum_{\langle i,j \rangle} \left[ (n_{u,j}+1)P(\ldots,n_{u,i}-1, n_{u,j}+1, \ldots, t) - n_{u,i}P \right]\nonumber \\
&+& \frac{\lambda}{2} \sum_{i} \left[(n_{v,i}-1)(n_{v,i}-2)(n_{u,i} +1)P(\ldots,n_{v,i}-1,\ldots, n_{u,i}+1, \ldots,t) -n_{v,i}(n_{v,i}-1)P \right] \nonumber \\
&+& \mu \sum_i  \left[ (n_{v,i}+1)P(\ldots, n_{v,i}+1, \ldots, t) - n_{v,i}P \right] + \nu \sum_i  \left[ (n_{u,i}+1)P(\ldots, n_{u,i}+1, \ldots, t) - n_{u,i}P \right] \nonumber \\
&+& f \sum_i  \left[ P(\ldots, n_{v,i}+1, \ldots, t) - P \right],
\ea
\end{widetext}
where $l$ is the characteristic length of the cell and $\langle \ldots \rangle$ denotes the sum over nearest neighbors.

The master equation~\eqref{master2} along with the sextic interaction shown in Fig.~\ref{fig:fig1} lends itself well to a many body description \cite{Doi76}, accomplished by the  introduction of an occupation number algebra with annihilation/creation operators
$\hat a_i, \hat a_i^\dag$ for $v$ and $\hat b_i, \hat b_i^\dag$ for $u$ at each site $i$. These operators obey the Bosonic commutation relations
\ba
\left[\hat a_i, \hat a^\dag _j \right] &=& \delta_{ij}, \quad \left[\hat b_i, \hat b^\dag _j \right]  = \delta_{ij}, \nonumber \\
\left[\hat a_i,\hat a _j \right] &=& 0, \quad~\left[\hat a_i^\dag,\hat a^\dag _j \right] =0,
\label{commute}
\ea
and define the occupation number operators $\hat n_{i,v} = \hat a_i ^{\dag} \hat a_i$ and $\hat n_{i,u} = \hat b_i ^{\dag} b_i$ satisfying the following eigenvalue equations:
\begin{equation}
\hat n_{i, v} | n _{i, v}\rangle = n_{i, v} | n _{i, v}\rangle,  \quad \hat n_{i, u} | n _{i, u}\rangle = n_{i, u} | n _{i, u}\rangle.  
\label{eigen}
\end{equation}
We next construct the state vector
\ba  \label{wavefunction}
|\Psi(t) \rangle =\sum_{\mathbf{n_v, n_u}}&& P(\mathbf{n_v, n_u} ,t) \nonumber \\
&& \times  \prod_i (\hat a^\dag_i)^{n_v ^i} (\hat b^\dag_i)^{n_u^i} |0 \rangle,
\ea
which upon differentiating with respect to time $t$, can be written in the suggestive form
\bq  \label{Schrodeq}
-\frac{\partial |\Psi(t) \rangle }{ \partial t} = H [\mathbf {\hat a ^{\dag}},\mathbf{\hat a},\mathbf{\hat b ^{\dag}},\mathbf{b}]|\psi_1(t) \rangle,
\eq
resembling the Schr\"odinger equation. Finally, taking  the time derivative of Eq. (\ref{wavefunction}) and comparing terms with the Hamiltonian in~\eqref{Schrodeq} we make the identification
\ba  \label{Hsym}
H &=& \frac{D _v}{l^2}  \sum_{\langle i,j \rangle}  (\hat a_i^\dag - \hat a_j^\dag)( \hat a_i- \hat a_j) + \mu \sum_i ( \hat a_i^\dag-1) \hat a_i   \nonumber \\
&&  + \frac{D _u}{l^2}  \sum_{\langle i,j \rangle}   ( \hat b_i^\dag-  \hat b_i^\dag) 
( \hat b_i- \hat b_j)  + \nu \sum_i( \hat b_i^\dag-1) \hat b_i \nonumber \\
&& -\frac{\lambda}{2}  \sum_i  \bigl[ \hat a_i^{\dag3} - \hat a_i^{\dag2} \hat b^{\dag} _i \bigr] \hat a_i^2  \hat b_i \nonumber \\
&& - f \sum_i(\hat b^\dag_i-1).
\ea

\subsection{Broken $U(1)$ symmetry and Doi shift}

\noindent The Hamiltonian in~\eqref{Hsym} has the property, 
\bq 
H[\mathbf{\hat a ^{\dag} = 1}, \mathbf{\hat a}, \mathbf {\hat b^{\dag} = 1}, \mathbf{ \hat b}]= 0,
\eq
implying that $H$ is identically zero at $ \mathbf{\hat a^{\dag}}, \mathbf {\hat b^{\dag} = 1}$.  
Furthermore, apart from terms involving $f,\mu,\nu$, we notice that $H$ is invariant under the $U(1)$ symmetry (representing particle number conservation) thus,
\ba
  a_i \rightarrow e^{i \theta} a_i  &&\quad a_i^\dag \rightarrow e^{-i \theta}  a_i^\dag, \nonumber \\
  b_i \rightarrow e^{i \theta} b_i  &&\quad b_i^\dag \rightarrow e^{-i \theta} b_i^\dag.
  \label{u1}
\ea
This symmetry is \emph{explicitly} broken by the dissipative and source terms. Nonetheless, the symmetric theory is key to identifying the composite structures that operate at short spatiotemporal scales and whose interactions preserve number conservation. Note that the vacuum state is defined by
\begin{equation}
\mathbf{\hat a | 0 \rangle} = 0, \quad \mathbf{\langle 0| \hat a^{\dag}} = 0.
\label{vacuum}
\end{equation}
Operation on the vacuum from either the left or the right does \emph{not} lead to the vanishing of $H$. Additionally, $H$ is non-hermitian, so its identification with a probability density (as in quantum mechanics) cannot be made.  Finally, translating~\eqref{Schrodeq} to Langevin dynamics (which is the goal of this paper) does \emph{not} lead to the correct kinetic terms in Eq.~\eqref{NGS}. 

The way to solve this problem was first proposed by Doi~\cite{Doi76, Grassberger_Scheunert}. He noticed that if the creation operators $\hat a_i ^{\dag}, \hat b_i^{\dag}$ are shifted by a single unit for each site, then the \emph{shifted} Hamiltonian $\tilde H$ has the desired properties.  To do this we define a displacement operator
 \bq
 D(\boldsymbol {\alpha}) = \textrm{exp}\left( \boldsymbol{\alpha . \hat a}\right),
 \label{dop}
 \eq
  which through the commutation relations~\eqref{commute}, leads to the following similarity transformation
\ba
D(\mathbf{1})~ F[\mathbf{\hat a^{\dag}, \hat a }]~D^{-1}(\mathbf{1}) &=&  F \mathbf{[ \hat a^{\dag} + 1, \hat a} ],
\label{similarity} 
\ea
where $F[\mathbf{\hat a^{\dag}, \hat a}]$ is an arbitrary function of the creation/annihilation operators.
Applying this transformation  to Eq.~\eqref{Hsym}, we now have,
\ba
\tilde{H} &=& \frac{D _v}{l^2}  \sum_{\langle i,j \rangle}  (\hat a_i^\dag - \hat a_j^\dag)( \hat a_i- \hat a_j) + \mu \sum_i \hat a_i^\dag \hat a_i   \nonumber \\
&&  + \frac{D _u}{l^2}  \sum_{\langle i,j \rangle}   (b_i^\dag - b_j^\dag) 
( \hat b_i- \hat b_j)  + \nu \sum_i \hat b_i^\dag\hat b_i \nonumber \\
&& -\frac{\lambda}{2}  \sum_i  \left(\hat a^\dag -  \hat b^\dag \right)   \left(1+ \hat a^{\dag2} + 2 \hat a ^{\dag}\right) \hat a^2 \hat b \nonumber \\
&& - f \sum_i\hat b^\dag_i,
\label{Hdoi}
\ea
where $\tilde{H}[\mathbf{\hat a ^{\dag}}, \mathbf{\hat a}, \mathbf {\hat b}, \mathbf{b}]  =   H[\mathbf{\hat a ^{\dag} + 1}, \mathbf{\hat a}, \mathbf {\hat b +1}, \mathbf{b}]$ is also known as the Doi-shifted Hamiltonian. Multiplying Eq.~\eqref{Schrodeq} on the left by the displacement operator $D(1)$ now yields,
\bq
-\frac{\partial |\tilde \Psi(t) \rangle }{ \partial t} = \tilde H [\mathbf {\hat a ^{\dag}},\mathbf{\hat a},\mathbf{\hat b ^{\dag}},\mathbf{b}]|\tilde \psi_1(t) \rangle,
\label{Schrodeq2}
\eq
where $\tilde\Psi(t) = D(1) \Psi(t)$. It is now straightforward to see that $\partial_t \langle 0 | \tilde \Psi(t) \rangle = 0$ and $\langle 0 | \tilde \Psi(t) \rangle$ is constant. In terms of an initial state vector $| \tilde \Psi(0)\rangle$, Eq.~\eqref{Schrodeq2} has the solution,
\bq
| \tilde \Psi(t) \rangle  = e^{-\tilde H t} |\tilde \Psi(0) \rangle,
\eq
which, as one can check, is completely equivalent to solving the master equation~\eqref{master2}.

We see, therefore, that the Doi-shifted Hamiltonian, although non-hermitian, conserves probability and still allows us to connect $\tilde \Psi(t)$ with a probability density. 
Furthermore, it is $\tilde H$ and not $H$ that recovers the correct chemical kinetics in~\eqref{NGS}. Interestingly, after the shift we are now at a different
minimum (corresponding to a broken symmetry vacuum) where it is the \emph{dissipative} terms that are now invariant to the symmetry transformations in~\eqref{u1}. 

\section{Path Integral formalism}
Having defined the space, the appropriate wave function and the correct Hamiltonian, we next seek to evaluate the operator $ e^{- \tilde H t} $ using the path integral formulation. This will enable us to uncover potential intermediate states present in a particular scale, that are however implicitly integrated out at different scales. 

Following the standard procedure for obtaining the coherent state path integral~\cite{NO98, THVL_2005, Tauber_2007} to the GS system, letting the coherent state $\phi_v$  (related to the operator $a$)  represent $v$ and $\phi_u$ (related to the operator $b$)  represent $u$ we obtain
\bq
e^{-\tilde Ht} = \int \calD \phi_v \calD\phi_v^\star \calD \phi_u \calD\phi_u^\star e^{- S[\phi_v,\phi_v^\star,\phi_u,\phi_u^\star]}, \label{pathint}
\eq
 where the action $S$ for the unshifted theory is given by
\ba
S = && \int dx \int_0^\tau dt \bigl[ \phi_v^\star \partial_t \phi_v + D_v \nabla \phi_v^\star \nabla \phi_v  +\phi_u^\star \partial_t \phi_u \nonumber \\
&&+ D_u \nabla \phi_u^\star \nabla \phi_u+ \mu (\phi_v^\star-1)  \phi_v + \nu (\phi_u^\star -1) \phi_u \nonumber \\
&& - f (\phi_u^\star -1)  
 - \frac {\lambda}{2}(\phi_v^\star-\phi_u^\star)\phi_v^{\star 2}  \phi_v^2 \phi_u  \bigr] ,
 \label{actionunshifted}
\ea
and for the Doi shifted theory we have instead,
\ba
S = && \int dx \int_0^\tau dt \bigl[ \phi_v^\star \partial_t \phi_v + D_v \nabla \phi_v^\star \nabla \phi_v  +\phi_u^\star \partial_t \phi_u \nonumber \\
&&+ D_u \nabla \phi_u^\star \nabla \phi_u+ \mu \phi_v^\star \phi_v + \nu \phi_u^\star \phi_u - f \phi_u^\star   \nonumber \\
&& 
 - \frac {\lambda}{2} (\phi_v^\star-\phi_u^\star)(1+2 \phi_v^\star +  \phi_v^{\star 2}) \phi_v^2 \phi_u  \bigr],
 \label{actionraw}
\ea
with $\phi_{v,u}$ being independent complex classical fields. 

By adding external sources to the action in the form~\eqref{pathint} one can calculate the connected Green's functions using perturbative expansions in the standard way. 
We are interested however in determining the Langevin dynamics associated with Eq.~\eqref{master2}. The equations of motion can be obtained by differentiating the action with respect to the starred fields.  We notice that the action given by ~ Eq. \eqref{actionraw} has terms linear, quadratic and cubic in the starred variables.  The terms linear in the starred variables lead to the classical reaction diffusion equations  in~Eq.\eqref{NGS}. The terms quadratic in the starred variables coming from the last line in  Eq. \eqref{actionraw}
(considered by \cite{ZHM06}) represent corrections to Eq.~\eqref{NGS} and are interpreted as \emph{internal} noise terms that are generated by the microscopic dynamics.  This interpretation is originally due to Feynman and Vernon \cite{FV63}  and since then has become a standard procedure \cite{Cardy99,Cal-Hu2008}.  
There are also terms cubic in the starred variables  $ - \frac {\lambda}{2}  (2 \phi_{v} ^{\star 3} -\phi_{v} ^{\star 2} \phi_{u}^{\star})\phi_v^2 \phi_u$.These terms are of particular physical significance as the term involving $\phi_v^{\star 3}$ represents inelastic scattering---corresponding to the chemical reaction $U+2V \rightarrow 3V$---while the term quadratic in $\phi_v ^{\star}$ represents the elastic scattering $U+2V \rightarrow U + 2V$ that maintains particle number conservation. To proceed to a Langevin description, however, we need to interpret these as noise terms, and it is therefore \emph{necessary} to convert them to terms quadratic in the starred variables by now defining the composite fields  
$\psi_1 =\phi_v^2$, $\psi_1^\star = \phi_v^{\star2}$.  Introducing these composites through Lagrange multiplier fields 
utilizing a representation of  the unit operator we have,
\ba
1&=&  \int \calD \psi_1 \calD \psi_1^\star \delta(\psi_1-\phi_v^2) \delta(\psi_1^\star - \phi_v^{\star2}) \nonumber \\
&=& \int \calD \chi^\star \calD \chi d\psi_1 d \psi_1^ \star \exp  \bigl[ -\chi^\star (\psi_1-\phi_v^2) \nonumber \\
&&- \chi(\psi_1^\star - \phi_v^{\star2}) \bigr].
\ea
At the expense of introducing the two composite fields $\psi_1, \psi_1^\star$ and the Lagrange multiplier fields $\chi, \chi^\star$ we can
write all the terms in the action as either linear or quadratic in the $\star$ field variables.  The action can now be written as
\begin{widetext}
\ba
S &=& \int dx \int_0^\tau dt \biggl[ \phi_v^\star \partial_t \phi_v + D_v \nabla \phi_v^\star \nabla \phi_v  +\phi_u^\star \partial_t \phi_u + D_v \nabla \phi_u^\star \nabla \phi_u
+ \mu \phi_v^\star \phi_v + \nu \phi_u^\star \phi_u - f \phi_u^\star
\nonumber \\
&&   - \frac {\lambda}{2} \psi_1 \phi_u \left[(\phi_v^\star-\phi_u^\star)+2 \psi_1^\star- 2 \phi_v^\star \phi_u^\star + \psi_1^\star(\phi_v^\star-\phi_u^\star) \right]
+\chi^\star (\psi_1-\phi_v^2)+\chi(\psi_1^\star - \phi_v^{\star2})  \biggr],
\label{actionpsi}
\ea
\end{widetext}
where the linear and quadratic terms are
\ba
S_{l} &=&\int dx [ \phi_v^\star \left(\partial_t - D_v \nabla^2  + \mu\right)\phi_v  - \sigma \phi_v^\star    - f \phi_u^\star +  \sigma \phi_u^\star  \nonumber \\
&& + \phi_u^\star \left(\partial_t - D_u \nabla^2  + \nu \right)\phi_u + \chi^\star (\psi_1-\phi_v^2)\nonumber \\
&& + \psi_1^\star ( \chi-2 \sigma), 
\ea
and
\bq
\tilde{S_q}=  -\chi \phi_v^\star \phi_v^\star\ - \sigma \left[  \psi_1^\star(\phi_v^\star-\phi_u^\star) - 2 \phi_v^\star \phi_u^\star \right], \label{sq}
\eq
with $\sigma = \frac{\lambda}{2}\phi_u \psi_1$.  

This can be written more compactly if we introduce the notation 
\ba
\tilde \phi (x) &=& \left(\phi_v, \phi_u ,  \psi_1 , \chi  \right) \nonumber \\
\tilde \phi^{\star}(x) &=& \left(\phi_v  ^{\star}  , \phi_u  ^{\star},  \psi_1 ^{\star} , \chi^\star \right),
\label{compact}
\ea
and source terms for $ \tilde \phi$ and $\tilde \phi^\star$. Doing so, we can write down the 
generating functional for the density correlation functions thus, 
\ba
Z[\tilde J,\tilde J^{\star}] = \int \calD  \tilde \phi \calD  \tilde \phi^\star  \exp  \bigl[- \{S_{l}[\tilde \phi, \tilde \phi^\star] + S_{q}[\tilde \phi, \tilde \phi^\star]\} \nonumber \\
+  \int dx  ( \tilde J_i  \tilde \phi_i + \tilde J _i ^\star \tilde \phi^\star_i) \bigr],
\label{genfunc}
\ea 
where we have absorbed the source terms $f$ and $\sigma$ into $\tilde J_i$.  The linear and quadratic pieces of the action are now represented in the compact form
\ba
S_{l} &&= \int dx dy    \left[\tilde \phi_i^{\star}(x) G_{ij}^{-1}(x,y) \tilde \phi_j(y) \right]  \nonumber \\
S_{q}&& = - \frac{1}{2} \int dx dy \left[\tilde \phi^\star_i(x)   D_{ij}(x,y) \tilde \phi^\star_j(y)\right],\nonumber \\  
\label{gen}
\ea
where the explicit form of the matrix $\mathbf{D}$ (which will become important shortly) is 
\bq
\mathbf{D} = 
\begin{bmatrix}
2\chi & -2\sigma & \sigma & 0 \\
-2\sigma & 0 &- \sigma & 0  \\
\sigma &- \sigma & 0 & 0  \\
0& 0 & 0 & 0 \\
\end{bmatrix}.
\label{Dij1}
\eq
Note that both matrices $\mathbf G^{-1}$ and $\mathbf{D}$ are functions of the fields $\phi_{v,u}$.
\section{Langevin Equations}

We notice that $S_q$ in Eq. \eqref{sq} is almost in the correct form to be interpreted as coming from a noise source added to the chemical kinetics equations \ref{NGS} for the GS model. The one difficulty is that the term $D(x,y)$ is itself a function of the fields, and this leads to yet another functional determinant in the action when going from a Langevin equation to a path integral.  To avoid this determinant we have to rewrite the multiplicative noise in terms of white noise or, equivalently, factorize  $D(x, x')$ using  the Cholesky decomposition and change bases to the white noise basis and thus remove the determinant.
First we introduce the  noise field $\eta(x)$ by employing the Gaussian identity  

\begin{widetext}
\bq
\sqrt{|\mathbf{D}|}\mathrm{exp}\left[ \frac{1}{2} \int dx dy \left[   \tilde \phi^\star_i(x)   D_{ij} (x,y) \tilde \phi^\star_ j(y)  \right] \right] =  \int {\cal D} \eta \exp \left[-\frac{1}{2} \int dx dy \left(\eta_i(x)  D_{ij}^{-1}(x,y)\eta_j(y) \right)+ \int dx \tilde \phi^\star_i(x) \eta_i(x) \right],
\eq
\end{widetext}
where $|\ldots|$ refers to the matrix-determinant. The probability distribution function for the noise is then,
\bq
P[\eta] = N \exp \left[-\frac{1}{2} \int dx dy \left[ \eta_i(x) D_{ij}^{-1}(x,y) \eta_j(y)\right]\right], 
\label{peta}
\eq
and the correlation functions are  $\langle \eta_i (x) \eta_j(y) \rangle = D_{ij} (x,y)$, with $D$ given by Eq. \eqref{Dij1}.

Next we eliminate the field-dependent determinant.
Since, $\eta_{\chi} = 0$, we need only consider the $3\times3$ sub-matrix of~\eqref{Dij1} which can be factored via the Cholesky decomposition thus,
\bq
\mathbf{D} = 
\begin{bmatrix}
2\chi & -2\sigma & \sigma  \\
-2\sigma & 0 &- \sigma   \\
\sigma &- \sigma & 0   \\
\end{bmatrix}
 = \mathbf{M^{T}M}.
 \label{cholesky}
\eq
With this factorization the exponent in~\eqref{peta} is,
\bq
\mathbf {\eta D^{-1} \eta = \eta^{T} M^{-1} (M^T)^{-1} \eta} = \theta^{T} \theta,\eq
with $\eta = \mathbf{M^{T}} \theta$, where $\theta$ is a white noise. Noting that $\sqrt{\mathbf{|D|}} = |\mathbf{M}|$,  the noise probability function can be re-written,
\ba
P[\theta] &=& P[\eta] \left |\frac{\delta \eta}{\delta \theta}\right |
=   \exp\left[-\frac{1}{2} \theta(x)^{T}\theta(x)\right],
\ea
where the white noise correlation functions are $\langle \theta_i (x) \rangle = 0$ and $\langle \theta_i(x) \theta_j (y) \rangle = \delta_{ij} \delta (x,y)$.
Following this the quadratic piece of the action is now,

\bq
S_q =  \int dx \left[ {\tilde \phi^\star}_i(x) G_{ij}^{-1}  (x,y) \tilde \phi_j(y) - \tilde \phi_i^{\star} (x) \eta_i (x)  \right],
\eq
and the Langevin equations for the GS chemical reactions are derived from
\bq
\frac{\delta S}{\delta \tilde \phi^{\star}_i(x)} = \int dx~G^{-1}_{ij} (x,y) \tilde \phi_j(y) - \eta_i(x) = 0.
\eq
These are
\ba  \label{NGS2} 
\frac{\partial \phi_v}{\partial t} && = \lambda  \phi_u \phi_v^2  - \mu \phi_v + D_v \nabla^2 \phi_v + \eta_v(x,t), \nonumber \\
\frac{\partial \phi_u}{\partial t} && =f - \lambda  \phi_u \phi_v^2 - \nu \phi_u + D_u \nabla^2 \phi_u + \eta_u(x,t), \nonumber \\
 \chi&& = \lambda  \phi_u \phi_v^2 + \eta_{\psi_1}.
\ea
The noise correlation functions are explicitly, 
\ba \label{noise3}
&& \langle \eta_v(x,t) \eta_v(y,t') \rangle = 2 \chi \delta (x-y) \delta (t-t') \nonumber \\
&& \langle \eta_u(x,t) \eta_v(y,t') \rangle = - 2\sigma  \delta (x-y) \delta (t-t') \nonumber \\
&& \langle \eta_v(x,t) \eta_{\psi_1} (y,t') \rangle = \sigma  \delta (x-y) \delta (t-t') \nonumber \\
&& \langle \eta_u(x,t) \eta_{\psi_1} (y,t') \rangle =- \sigma   \delta (x-y) \delta (t-t') \nonumber \\
&& \langle \eta_u(x,t) \eta_u(y,t') \rangle  =  \langle \eta_{\psi_1} (x,t) \eta_{\psi_1} (y,t') \rangle  = 0.
 \ea
A few comments about these correlation functions are in order.  The amplitudes of the correlation functions involving the combination of $\phi_u$ and $\phi_v$  are negative, indicating that the concentrations $u$ and $v$ are anti correlated with the noise amplitude being complex. This is in fact similar to what happens in the case of annihilation of a molecule at a single site as is  discussed in \cite{Cardy99}. More remarkable is the following:  the amplitude of the auto-correlation function for $\eta_v$  depends on the composite field $\chi$, which in turn depends on a noise source as well. This indicates the presence of some form of stochastic feedback. This source is   correlated with $v$ but anti-correlated with $u$. This noise effects only higher order  (two loop) processes in perturbation theory but is crucial for understanding the renormalization of the coupling constant.  

As mentioned before, we can rewrite the multiplicative noises as a combination of white noise factors, via the Cholesky decomposition in Eq.~\eqref{cholesky}. Incidentally, the factorization is not unique. However, one simple realization yields 

\bq \label{factor} 
\mathbf{M^T} = \frac{1}{\sqrt{2 \chi}}
\begin{bmatrix}
2 \chi & 0&  0 \\
-2 \sigma & -2 i  \sigma & 0\\
\sigma  & i (\sigma- \chi)  & i  \sqrt{\chi (2 \sigma - \chi)} 
\end{bmatrix},
\eq
in which case the noises are explicitly,
\ba
\eta_v &&=  \sqrt{2 \chi} \theta_1, \quad \eta_u  = -\frac{2 \sigma}{\sqrt{2 \chi}} (\theta_1 + i \theta_2),\nonumber \\
\eta_{\psi_1} && =\frac{1}{\sqrt{2 \chi}} \left[ \sigma  \theta_1 + i (\sigma - \chi) \theta_2+ i \sqrt{\chi (2 \sigma - \chi)} ~\theta_3 \right].  \nonumber \\
\ea
Numerical simulations of these systems of equations can be implemented by a separation of the real and imaginary parts of the fields. As the fields $\phi_{v,u}$ must in the end correspond to real physical densities, the noise averaged imaginary parts of the fields should vanish, while the real component $\langle \phi_{v,u}\rangle_\eta$ will correspond to the positive concentrations $v,u$ (see~\cite{ZHM06}).

\begin{figure*}[t]
\centering 
\includegraphics[width=0.75\textwidth]{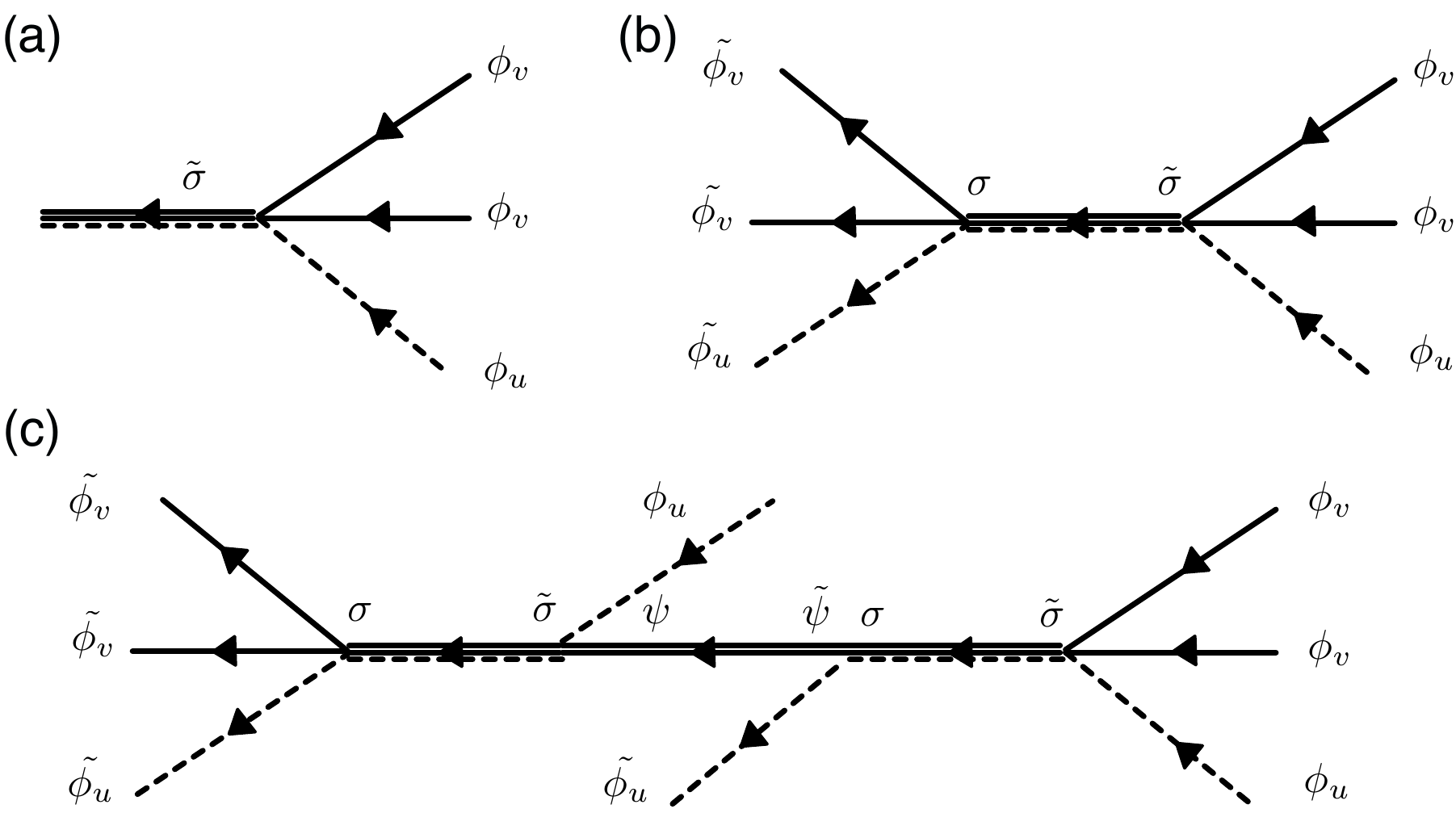}
\caption{Decomposing the scattering shown in Fig.~\ref{fig:fig1} in terms of composite field correlation functions .  (a) The ``bare" vertex in the unbroken version of the theory~\eqref{actionunshifted}, where the composite field propagator for  $\sigma$ is a sum of two loop graphs made from 2 $\phi_v$ and one $\phi_u$ fields. (b) In this language the elastic scattering $U+2V\rightarrow U + 2V$ can be thought of as going through an intermediate $\sigma$ resonance.  (c) The fluctuation induced process $2U+2V \rightarrow 2U+2V$ ($4\rightarrow 4$ scattering )where the composite $\sigma$ shakes of a $\phi_u$ to form a composite field correlation function for $\psi_1$, a one loop graph made of 2 $\phi_v$ propagators. This then recombines with a $\phi_u$ to reform $\sigma$ which then decomposes into the original fields. Similar diagrams exist for the inelastic scattering and all higher order $n \rightarrow n$ processes.}
\label{fig:fig2}
\end{figure*}

\section{Fluctuation induced processes}
While the Doi-shifted version of the action \eqref{actionraw} generates the classical Gray-Scott  ``chemistry'' with the appropriate internal noise terms, it is easier to determine the graphical structure of the interactions induced by the fluctuations from the unshifted form of the action~\eqref{actionunshifted}.  In this case one has the two basic interactions,  the inelastic scattering corresponding to $\phi_u \phi_v^2 \rightarrow  \phi_v^3$  {\it and} the elastic scattering $\phi_u \phi_v^2 \rightarrow \phi_u \phi_v^2$.   Due to the presence of directionality in the interactions, loops get generated from multiple elastic re-scatterings, although there are no one particle irreducible corrections to the bare propagators for $\phi_{u,v}$. Instead, this process renormalizes the local bare scattering graph leading to a geometric series of two loop graphs identical to that found for the annihilation reaction for a chemical $A$, where $3A \rightarrow 2A$ as discussed in detail in~\cite{THVL_2005}.  

As in the broken-symmetry version~\eqref{actionraw} we can introduce the composite field $\sigma = \lambda \phi_u \phi_v^2$, and think of the scattering as going through a $\sigma$ resonance. The sum of the two loop graphs  can then be interpreted as the ``propagator" for the composite field $\sigma$.  
Because of the $U(1)$ symmetry, the interactions in the GS model must preserve particle number conservation.  Thus one induces from the fluctuations $n \rightarrow n$ particle interactions starting from the $n \ge 4$ loop level.

In terms of the correlation functions for the composite field $\psi_1=\phi_v^2$, the $n \rightarrow n$ interactions can also be summed exactly in the unbroken theory and written in terms of ``tree" diagrams in the original propagators for $\phi_u$ and $\phi_v$ combined with the composite propagators for $ \psi_1, \sigma$.   This is illustrated for the  $4 \rightarrow 4$ scattering process  $2U + 2 V \rightarrow 2U + 2 V$ in Fig.~\ref{fig:fig2}.  In fact by identifying another composite $\psi_2 = \phi_u \phi_v$ we can also generate reactions of the form $U  + 3V \rightarrow U + 3V$ whose structure is analogous to that shown in Fig.~\ref{fig:fig2}(c). These graphs (and the ones in the broken-symmetry version) can also be generated in perturbation theory by solving the Langevin equations as a power series in $\lambda$ and then taking the averages over the noise correlation functions.
This will be discussed in detail in a future paper.

\section{Conclusions}
In this paper we have discussed in detail the derivation of the Gray-Scott model starting from the formulation of its master equation and eventually obtaining the corresponding Langevin equation that includes internal noise terms. In the process of carrying this out we have uncovered a very rich structure underlying the model that suggests the presence of  a hierarchy of scales~\cite{prlus}.  

The master equation for the GS model has previously been studied in~\cite{ZHM06} where the path integral for the evolution operator of the many-body wave function was derived, however only a part of the nonlocal interactions in the Hamiltonian was converted into an equivalent system of stochastic differential equations, thereby violating the conservation of probability (unitarity). The overall conservation of probability requires the introduction of composite fields which in turn are stochastic in nature. Not surprisingly these fields can be interpreted as the harbingers of additional hierarchical structures. 

By introducing these composite fields representing degrees of freedom  faster than those in Eq.~\eqref{NGS}, we obtained an alternative Lagrangian which turns out to be quadratic instead of cubic in the conjugate starred fields. Following this, we were able to show that the correlation functions for the model could be obtained by solving a system of Langevin equations whose noise is specified by a particular Gaussian distribution for its correlation functions. We find that the cross-correlated noise (i.e involving $ \phi_u$ and $ \phi_v$) is actually anti-correlated.  This is not surprising: in the Gray-Scott model, by design, an increase in the concentration of $U$ corresponds to a decrease in the concentration of $V$ and vice versa. The anti-correlation is merely a stochastic manifestation of this phenomenon.  On the other hand, the amplitude of the noise correlation function involving only $\phi_v$ is proportional to a composite field $\sigma$ plus a noise term.  The field $\sigma$ is the composite of $\phi_u \phi_v^2$, which suggests that this three body resonance plays an important role in the internal dynamics.  Indeed in the unbroken theory, the elastic scattering of $U+2V \rightarrow U + 2V$ can be interpreted as proceeding through the formation of a $\sigma$-resonance alone.

In the process of determining the Langevin equation, we have discussed the importance of the explicit breaking of a global $U(1)$ symmetry in the Hamiltonian.  This requires a shift to a new vacuum and induces the correct mean field level Gray-Scott equations.  In some ways, this is reminiscent of the spontaneous breaking of a global $U(1)$ symmetry in an interacting Bose gas.  There, the leading order mean field theory yielding the classical behavior of the condensate is obtained by  expanding around the broken symmetry minimum, eventually leading to the Gross-Pitaevskii equation for the condensate~\cite{Gross61,Pitaevskii61}.
 
We have also  derived the general structure of the path integral for the correlation functions of the model rewritten in terms of the composite fields. When reformulating the model in terms of these composite fields, the structure of the interaction takes the suggestive form depicted in Fig.~\ref{fig:fig2}.  The equations we have derived will enable us to determine the nature of the critical phenomena that lead to the observed large scale behavior mentioned in the Introduction.
 By studying the properties of the Langevin equations, or equivalently their associated Schwinger--Dyson equations for the correlation functions, we will also be able to determine the nature of the critical phenomena that lead to the observed large scale behavior mentioned in the Introduction.  These include phenomena such as formation of domains, their self-replication and co-operative effects and will be explored using the dynamical renormalization group formalism applied to Eqns.~\eqref{NGS2}~and~\eqref{noise3} (or equivalently ~\eqref{genfunc}) in subsequent work.

Finally, we would like to point out that the methodology of introducing composite field operators transcends the application presented here. Indeed, this is applicable whenever one would like to formulate a Langevin description of reactions where there are three (or more) chemical agents participating in a reaction---the important class of Branching and Annihilating Random Walks~\cite{THVL_2005}, or in quantum field theory interactions that are quartic or higher order, such as calculating the kinetics of condensates in Bose gases~\cite{Kamenev11}). The Langevin description enables a simple approach for numerical computation of correlation functions.
It also can be used to directly obtain an ``Effective Potential" that depends only on the fields $\phi$ and not the conjugate fields \cite{Hochberg_etal_1999}.  Additionally, in certain systems, one can use the composite field technique in conjunction with the renormalization group to exhaustively determine the subcomponents of a system, should they be present.

\begin{acknowledgments}
The authors would like to thank John Dawson, Uwe T\"{a}uber and David Hochberg for valuable discussions and Repsol S.A. for funding the research reported here.
F.C. would like to dedicate this paper to Dr. Prem Shekar who successfully performed open heart surgery on him. 
\end{acknowledgments}


\begin{thebibliography}{10}

\bibitem{Mandelbrot}
B. Mandelbrot, {\em The Fractal Geometry of Nature} (W. H. Freeman and Co., New
  York, 1982).

\bibitem{PerBak}
P. Bak, {\em How Nature Works: The Science of Self-Organised Criticality}
  (Copernicus Press, New York, 1996).

\bibitem{Holland}
J.~H. Holland, {\em Emergence: From Chaos to Order} (Perseus Publishing, New
  York, 1998).

\bibitem{prlus}
F. Cooper, G. Ghoshal, A. Pawling, and J. {P\'erez-Mercader}, Phys. Rev. Lett. {\bf
  111},  044101  (2013).

\bibitem{CG09}
M. Cross and H. Greenside, {\em Pattern Formation and Dynamics in
  Nonequilibrium Systems} (Cambridge Univ. Press, Cambridge, 2009).

\bibitem{Walgraef97}
D. Walgraef, {\em Spatio-Temporal Pattern Formation} (Springer, New York,
  1997).

\bibitem{GS83}
P. Gray and S.~K. Scott, Chem. Eng. Sci. {\bf 38},  29  (1983).

\bibitem{Pearson93}
J.~E. Pearson, Science {\bf 261},  189  (1993).

\bibitem{LHMPM03}
F. Lesmes, D. Hochberg, F. Mor\'an, and J. {P\'erez-Mercader}, Phys. Rev. Lett.
  {\bf 91},  238301  (2003).

\bibitem{HLMPM03}
D. Hochberg, F. Lesmes, F. Moran, and J. P\'erez-Mercader, Phys. Rev. E. {\bf
  68},  066114  (2003).

\bibitem{Kamenev11}
A. Kamenev, {\em Field Theory of Non-Equilibrium Systems} (Cambridge Univ.
  Press, Cambridge, 2011).

\bibitem{Altland}
A. Altland and B. Simons, {\em Condensed Matter Field Theory} (Cambridge Univ.
  Press, Cambridge UK, 2010).

\bibitem{VK83}
N.~G. {Van Kampen}, {\em Stochastic Processes in Physics and Chemistry}
  (North-Holland, Amsterdam, 1983).

\bibitem{Doi76}
M. Doi, J. Phys. A: Math. Gen. {\bf 9},  1465  (1976).

\bibitem{CJP74}
S. Coleman, R. Jackiw, and H.~D. Politzer, Phys. Rev. D {\bf 10},  2491
  (1974).

\bibitem{dMRE93}
C.~A.~R. S\'a~de Melo, M. Randeria, and J.~R. Engelbrecht, Phys. Rev. Lett.
  {\bf 71},  3202  (1993).

\bibitem{CCMD10}
F. Cooper, C.-C. Chien, B. Mihaila, J. F. Dawson and E Timmermans, Phys. Rev. Lett. {\bf 105},  240402  (2010).

\bibitem{HZM_2005}
D. Hochberg, M.-P. Zorzano, and F. Moran, J. Chem. Phys. {\bf 122},  214701
  (2005).

\bibitem{Grassberger_Scheunert}
P. Grassberger and M. Scheunert, Fortschritte der Physik {\bf 28},  547
  (1980).

\bibitem{NO98}
J.~W. Negele and H. Orland, {\em Quantum Many-Particle Systems} (Perseus
  Publishing, Cambridge, 1998).

\bibitem{THVL_2005}
U.~C. T\"{a}uber, M. Howard, and P. Vollmayr-Lee, J. Phys. A: Math. Gen. {\bf
  38},  R79  (2005).

\bibitem{Tauber_2007}
U.~C. T\"{a}uber,  in {\em Aging and the Glass Transition}, Vol.~716 of {\em
  Springer Lecture Notes in Physics}, edited by M. Henkel, M. Pleimling, and R.
  Sanctuary (Springer-Verlag, Berlin, 2007), pp.\ 295--348.

\bibitem{ZHM06}
M.-P. Zorzano, D. Hochberg, and F. Moran, Phys. Rev. E {\bf 74},  057102
  (2006).

\bibitem{FV63}
R. Feynman and F. Vernon, Annals of Physics {\bf 24},  118   (1963).

\bibitem{Cardy99}
J. Cardy, Field Theory and Non-Equilibrium Statistical Mechanics, Lectures
  presented as part of the Troisieme Cycle de la Suisse Romande, (1999).

\bibitem{Cal-Hu2008}
E. Calzetta and B.-L. Hu, {\em Nonequilibrium Quantum Field Theory} (Cambridge
  Univ. Press, Cambridge UK, 2008).

\bibitem{Gross61}
E. Gross, Il Nuovo Cimento Series 10 {\bf 20},  454  (1961).

\bibitem{Pitaevskii61}
L.~P. Pitaevskii, Soviet Physics JETP-USSR {\bf 13},  451  (1961).

\bibitem{Hochberg_etal_1999}
D. Hochberg, C. Molina-Paris, J. {P\'erez-Mercader}, and M. Visser, Phys. Rev.
  E {\bf 60},  6343  (1999).

\bibitem{Amit}
D. Amit, {\em Field Theory: the Renormalization Group and Critical Phenomena}
  (World Scientific, Singapore, 1984).

\bibitem{Binney}
J. Binney, N.~J. Dowrick, A. Fisher, and M.~E.~J. Newman, {\em The Theory of
  Critical Phenomena} (Oxford University Press, Oxford, England, 1992).

\end{thebibliography}
\end{document}